# OA@MPS — a colourful view

Laurent Romary
Max Planck Digital Library

Abstract

The open access agenda of the Max Planck Society, initiator of the Berlin Declaration, envisions the support of both the green way and the golden way to open access. For the implementation of the green way the Max Planck Society through its newly established unit (Max Planck Digital Library) follows the idea of providing a centralized technical platform for publications and a local support for editorial issues. With regard to the golden way, the Max Planck Society fosters the development of open access publication models and experiments new publishing concepts like the Living Reviews journals.

Zusammenfassung

Die Open Access Agenda der Max-Planck-Gesellschaft, Initiator der Berliner Erklärung, sieht sowohl eine Unterstützung des grünes als auch des goldenen Weges zu Open Access vor. Zur Umsetzung des grünen Weges verfolgt die Max-Planck-Gesellschaft durch seine neu gegründete Einheit (Max Planck Digital Library) die Idee der Bereitstellung einer zentralen technischen Plattform für Publikationen und einer lokalen Unterstützung bei redaktionellen Fragen. Hinsichtlich des goldenen Weges unterstützt die Max-Planck-Gesellschaft die Entwicklung von Open Access Publikationsmodellen und erprobt neue Publikationskonzepte wie die Living Reviews Zeitschriften.

In 2003, the Max Planck Society has been the initiator of the Berlin Declaration, which expresses a global vision on open access (OA) to scientific knowledge, and is now quoted as a reference statement in any open access endeavours. Among these, the recent years have seen many initiatives intended to foster the archival of scientific publications in open repositories (green way to OA) as well as the definition of new business models (gold way) that would lead to the large-scale implementation of the Berlin Declaration principles. Still, there remains work to be done before scientists and the general public all over the world have at their disposal the wide compendium of research results in all forms of presentation. In this respect, the Max Planck Society wants to keep the agenda moving ahead by exploring how it may integrate open access activities related to traditional publications, new publishing models and dissemination of research data in one single vision. The so-called open access agenda of the Max Planck Society addresses those issues along various dimensions, namely:

- Scientific: going towards the definition of scientific collaborative environments that would implement the role of open access in a wider notion of eScience;
- Technical: identifying the need for integrated and sustainable platforms for the management of research assets;
- Editorial: defining the measures to be taken to help researchers adhere to the open access principles and make their results usable to a wide scientific community;
- Political: contributing to increase open access awareness and the stronger coordination of institutions worldwide.

Our perception is that many roads to open access are still to be developed and we would like that the scientific community will join forces to creatively implement the Berlin Declaration. In this context, this paper, while not try to be exhaustive as to the issue of open access, would like to exemplify the debate in the light of the specificities of the Max Planck Society.

## The Max Planck Society and the Max Planck Digital Library

The Max Planck Society, beyond its renowned scientific excellence, can, from the point of view of scientific information, be observed from two main perspectives:

- The MPS is a *multidisciplinary* research organisation, covering most scientific fields in natural and human sciences.
- It is organized as a network of highly *autonomous institutes*, which, once founded, have full liability to plan and deploy their activities according to their own research agendas.

This implies that generic and centralized solutions for scientific information management can only be devised in close articulation with the local activities carried out in institutes. In particular, one has to keep in mind the central role of the libraries attached to most of them, which by essence are closely related to the local research needs.

As a consequence, it has always been difficult to align the strong global visibility of the Max Planck Society in the domain of open access, as backed-up by strong scientific personalities and highly relevant local initiatives, and the difficulty to deploy a general access policy within the institutes. This is one of the reasons that has lead to the foundation of a unitary service unit dedicated to scientific information management and dissemination, the Max Planck Digital Library (MPDL).

The MPDL provides services to help the researchers in the Max Planck Society manage their scientific information workflow. Such services comprise the provision of actual content and technical solutions, but also by acting as a centre of competence and community facilitator in scientific information management.

Importantly as well, the MPDL is in charge of the strategic issues related to the wide dissemination of research results towards the scientific community, and in particular in contributing to the design and implementation of the MPS open access policy.

The activities of the MPDL can be outlined along the following lines:
- ⇨ Content provision: The MPDL is in charge of negotiating and providing access to digital content to the institutes. The selection of such content is to be made in strong collaborations with the institutes (mainly through their libraries);
- ⇨ Technological development: The MPDL focuses on providing technological platforms and tools as a complement to what is being locally implemented in the institutes. The MPDL is thus in charge, in collaboration to the FIZ Karlsruhe, of the eSciDoc project, a platform for the management of digital publications and research data;
- ⇨ Expertise provision: Beyond the two preceding core activities, it is important to act as an interlocutor towards institutes, in order to advise about the best standards, practices and technological state of the art and make sure that each new project related to digital information is at least aware of what has been done elsewhere in the MPS;
- ⇨ Strategic planning: The MPDL is part of the various decision processes within the MPS, whenever they comprise aspects related to scientific information management. This will ensure both a coherence of the decisions and a memory of the underlying rationales of the decision taking process;
- ⇨ Networking: The MPDL contributes in grouping together scientists, institutes or other stakeholders that have similar (or complementary) needs and activities in the Max Planck Society.

Still, as we will see in this paper, the MPDL should not be in charge of the curational activities related to the creation (digitization) or management (metadata) of data. This should be kept at institute level, even if the MPDL may provide support to the planning and setting up of such activities.

# OA@MPG

## *Archiving publications*

The mainstream view on open access, as defended by its core supporters, is to foster the depositing of scientific articles in a publication archive, so that, according to community of practices, legal possibilities and/or institutional backing, the full text of the paper can be made widely accessible on-line. The corresponding version of the paper can either be the author's initial draft (pre-print), the manuscript after peer review (post-print) or the publisher's version, associated of course with various levels of dissemination freedom.

As a matter of fact, there are quite a few reasons why we may want a have our scientific outputs be archived systematically within a repository, and, when having a closer look at them at them these are only partially related to the issue of open access. Still, they all aim at being beneficial to scientists and scientific institutions, since having a full coverage of one's production within a reliable repository provides a digital memory of research results, which is an essential tool for scientific activities at large. Beyond bringing immediate access to the full text, it also gives the possibility to produce multiple views on publications, which in turn can be used to generate publication lists, web pages, selections of most relevant publications (thematic selections or to provide compendia for assessment committees). It thus brings the capacity for an institution to have a whole photography of its outputs that may be used for strategic planning or bibliometrical analyses (in complement to commercial tools), and allows it to keep an archive of past activities when departments or institutes are closed. Of course, by providing also open access to part of the content, such a repository can become a strong instrument of dissemination.

Still, such a view on publication repositories can only make sense if strong technical and editorial support is provided to provide simple yet effective services, and also guaranty the quality and relevance of its content, in particular from the point of view of metadata. This in turn implies finding a good organisational scheme that optimizes the means puts on such an archive both from the viewpoints of reliability, cost-effectiveness and scientific proximity.

We thus defend an organisation based, on the one hand, upon a highly centralized technological framework, and on the other hand, on a local editorial support to scientist. The central deployment of the archive prevents a technological fragmentation whereby several IT groups are uselessly duplicating maintenance, updating, and sometimes development works. It also allows to provide good central services for issues like dissemination and long-term preservation, but above all to provide quick and responsive answers to users's needs in domains like, workspace management, usage statistics or full text search.

On the contrary, the editorial support, i.e. the validation and possible enrichment of the data deposited by the scientists has to remain as local as possible, and is probably the natural extension of the usual missions of the libraries. This digital curatorship has to be made effective in the context of a good knowledge of the research communities and understanding of the corresponding expectations. One of the main duties here is in particular to check and improve the quality of the affiliations associated to the published articles. Libraries are also the level at which information concerning the publication archive can be provided, and conversely where user feedback can be gathered up and brought to the technical side

All this can only make sense if the perspective is to achieve as wide a coverage as possible within one's publication archive. This is why the Max Planck Society finds it necessary to go beyond incentive measures towards publication archiving and issues a deposit mandate for all publications corresponding to work carried out in its institutes. This decision, which will be finalized in the fall of 2007[1] will put no specific constraint on the further visibility of archived documents, but will allow us to have a stable basis for further open access related activities.

## *Turning up publication models*

The commitment of the Max Planck Society in gold open access results from the analysis that it makes no sense to push the green way without accompanying the unavoidable changes in publication practices and the related business models. To our view, the core factors that will lead to a fruitful collaboration between research institutions and publishers can be outlined as follows:

- Copyright transfer should be left out of any such agreement, so that independently of the certification and/or dissemination service provided by the publisher, full liability is left to the author to issue new dissemination formats or variants that he/she feels necessary to propagate his/her results;
- The institution should have the capacity to mirror the final paper in its own archive. This is an essential aspect for providing reliable data in situation like assessment campaigns;
- A strong collaboration has to be carried out to normalize affiliations so that researchers corresponding to a given institution are presented in a coherent way. Independently of addresses appearing on printable papers, it is essential to work towards agreements that would lead, in the long run, to a full compatibility between metadata in publishers' databases, institutional archives, and consequently commercial bibliographical databases;
- Last but not least, transparent cost models should allow research institutions or universities to choose the level of service they may require from publishers, with the expectation that cost saving can become a natural, and shared trend.

These various constraints together with priorities set by researchers themselves within the Max Planck Society have thus led us to articulate our policy along three main action lines:

- Taking part in multi-organisation consortia working towards global switches from traditional subscription based models to full open access. The MPS has thus strongly supported and contributed to the establishment of the SCOAP3 proposal;
- Design collaborative framework with full open access journals and publishers, in particular when there is a strong back by scholarly groups. This is typically the case with Copernicus, which, with the support of the European Geoscience Union, offers probably at present the most transparent and scientifically motivated open access scheme;
- Avoid the fragmentation of our financial and decisional surrounding by rejecting paper-based open access scheme in favour of global negotiation with traditional publishers. The underlying objective for us is here to make the gold open access process as transparent and administratively innocuous as possible.

As a whole, the policy of us going Gold is not to contribute to the preservation of the existing publishing ecology, but above all to contribute to make this ecology evolve in the direction we think would provide better services and at a better price for our scientists.

## *Are research data concerned at all?*

There are quite a few reasons to consider that open access to research data will become in the future an essential, or even the main, component of an open access policy for a research institution like the Max Planck Society. Indeed, this is already an issue that has been put high on the agenda by several research communities such as astronomers, geneticians or researchers in the history of science, who have started to develop communities and infrastructures to provide a wide dissemination of their digital assets. Depending on the scientific domain, the underlying urge to archive and disseminate research data comes from the need to pool together primary sources, to compare results but also, in relation to traditional publications to provide means to supply the evidence behind asserted claims.

---

[1] And will be published on the OA information platform by that time

Still, whereas sharing research data is obviously a need, a lot of factors precludes us from adopting a blunt and global open access policy in this respect. First, there can be quite a few legal issues preventing wide dissemination, related either to copyright restrictions on the sources (e.g. contemporary documents, museographic data) or the relation to personal information (personal data, photographies, medical descriptions). There is also the difficulty, when one deals with complex data structures, not only to provide the data itself, but also the corresponding tools to have actual means to exploit them. As a whole, the only relevant strategy in this respect is to accompany scientific communities when expressing needs related to strong research needs.

From the point of view of the Max Planck Society, we both contribute to disseminate the technical experience of communities which have already developed complex environments for the management and dissemination of data, while offering technical support, through the MPDL, for newcomers, focusing on generic solutions that may bring more and more researchers to a better management of their digital production. As the data to be preserved are very heterogeneous among the various MPIs virtual groups of institutes should be formed to bring together those of similar demands and interests.

As an example the seven MPIs working in the field of astronomy can all take advantage of the activities currently undertaken under the auspices of the International Virtual Observatory Alliance (IVOA). The German Astrophysical Virtual Observatory project GAVO, initiated by the MPIs for Extraterrestrial Physics (MPE) and for Astrophysics (MPA), is representing the German astronomy community. Aim of the 16 national VObs collaborating within the IVOA is the development of standards to ensure interoperability of their highly distributed data-centres containing very heterogeneous data-sets (in particular with respect to registries, metadata, protocols for accessing images, spectra, catalogues, numerical simulations, and related literature). The concept is designed in such a generic way that it can be adapted by other communities or for other purposes.

In the long run of course, the MPS will have to consider also mandating the archival and, when possible, the wide dissemination of all data produced associated to the publication of a research result. Still, it appears that this cannot be achieve before we have a comprehensive view of the means to be deployed to achieve this objective. An essential component of such strategy is related to having strongly trained personnel in digital curation techniques that will accompany researchers in their management of research data.

### *New Publication Platforms, New Publication Models*

Whether Green or Gold the traditional views on open access are based on the assumption that publication vectors remain unchanged, i.e. in the form of fixed published articles in journals as resulting from a closed peer-review process. Still, it is probably our duty to see what the development of new technical means can bring to us and explore new forms of scientific communication that could be adopted by all or some research communities.

In a way, this is exactly what has led to the creation of Arxiv[2], with a community of scientists extending their natural trend to exchange drafts among themselves and using the internet infrastructure to do so in a simplified manner. They actually opened the way for generalizing such environments, whether through thematic or institutional archives. One can observe though that each scientific community has projected its own perception on how such archives could be used, with very few scientists actually disseminating pre-prints through this channel.

There is also quite some room for evolution in the domain of "traditional" publishing and we can take two examples related to the Max Planck Society to illustrate this.

First, the impact on new technologies upon journal publishing can facilitate the management of paper versioning. This is the case with the Living Reviews series[3], which, in scientific fields ranging from physics to the humanities, publish several journals dedicated to high level state of the art papers. These are completely open access publishing vectors, with a high focus on scientific quality. Regular revisions contribute to make the corresponding papers unavoidable reference materials in the corresponding fields.

As a second example, we can have a quick glance at the publishing model deployed in most of Copernicus[4] journals (many of which are undertaken under the auspices of the European Geoscience Union). In this case, the capacity of providing immediate online access to information is used to provide a open access peer review process. From the stage of submission to that of final publication, all papers, and above all the corresponding reviews, are freely accessible thus ensuring a kind of global controlling capacity for the scientific community. The model has resulted in a clear change in publication practices within these journals. Less paper submissions, better reviews, and higher acceptance rates while preserving, even increasing, the scientific impact show that we should not be reluctant in providing new types of scientific communication.

Finally, there is a strong demand from some communities to have access to publishing channels allowing them to get scientific recognition for the activity they conduct in the domain of research data. Already explored in communities like genomics, where short papers can be associated to the deposit of a genomic sequence in a database, it appears to be a necessary environment for disciplines whose core activity is to analyse primary sources or objects, such as linguistics, archaeology or history. This leads to the idea (aka "living sources") that real peer reviewed publishing environment must be implemented whereby researchers can deposit data sets, together with annotations and/or commentaries, that in turn they can quote as part of their actual research production. This is to our view an important dimension for the future development of infrastructures such as eSciDoc, if we want them to be accepted by scientists.

### *Improving awareness*

As one can see from this overview of the various issues at hand, open access is a highly complex issue, even more, if it is taken for granted independently from the scientific diversity as observed in the various institutes of the Max Planck Society. Since there is no global OA solution, we want also to defend the idea that an OA dissemination policy should not be based on education (or evangelization), but on the capacity to listen to the scientists' needs or worries with regards to communication of their scientific results. By doing so, we have already identified that their main expectations rely not so much on OA as a principle, but on the capacity of the corresponding infrastructures to provide reliable and effective research environments for preserving and handling their own information. This rather self-interested view on scientific information has then to be matched against more systemic views on community or institution interests, so that the idea of open access per se becomes a natural component of the scientists' ecology.

---

[2] http://arxiv.org
[3] http://www.livingreviews.org/
[4] http://www.copernicus.org/

## Joining efforts

Most of the elements presented in this paper are not specific to the Max Planck Society and could be taken up by any other research institution or university in the definition of its scientific information strategy. In particular, most of the technological developments, as well as editorial support policies, are likely to be implemented or defined by others simultaneously. As a consequence, it is essential to contemplate the various possibilities that one has to join efforts nationally and internationally to avoid duplicate works, but also contradictory actions towards similar interlocutors, whether scientists, publishers or decision makers.

In this respect, endeavours aiming at coordinating activities on publication archives (Driver[5]), research data management (Dariah[6]) or open access communication (OA information platform[7]) play an essential role in ensuring a better synergy between institutions, but also foster the development of new ideas in the field of open access. These are also places where we could probably implement the dual central-decentral strategy that we presented for the Max Planck Society.

As a final word of conclusion, we can say that in the long run, open access is a non-avoidable target. It is technologically feasible in principle, but above all, it is the only way to improve the quality and dissemination of research worldwide. Still, if we want this movement to be really useful for science, we have to consider how research organisations as well as individual scientists can go towards a coherent and efficient scheme for the wide dissemination of scientific results.

## Acknowledgments

This paper has been written on the basis of numerous discussions that have been held within the Max Planck Society. I am in particular most grateful to my colleagues in the sInfo steering committee and Max Planck Digital Library[8] for having brought so many complementary ideas in the debate. It has also benefited from the experience gained in the French research environment both at CNRS[9] and INRIA[10].

---

[5] http://www.driver-repository.eu/

[6] http://www.dariah.eu/

[7] http://open-access.net/

[8] For further contact about the Max Planck Society activities, contact the OA group: Christoph Bruch and Anja Lengenfelder.
See http://open-access.net/de/oa_informationen_der_maxplanckgesellschaft/

[9] http://openaccess.inist.fr/

[10] http://www.inria.org/publications/archiveouverte/